\documentclass[
reprint,                % Стиль как в окончательной версии статьи
superscriptaddress,     % Авторы с надстрочной нумерацией адресов
amsmath,                % Расширенная поддержка математики AMS
amssymb,                % Дополнительные математические символы
prab,                    % Стиль Physical Review A
]{revtex4-2}            % Класс документа REVTeX 4.2

\usepackage[utf8]{inputenc} % Поддержка UTF-8 (для кириллицы и спецсимволов)
\usepackage[T2A]{fontenc}   % Кодировка шрифтов для кириллицы
\usepackage[russian,english]{babel} % Поддержка двуязычного документа

\usepackage{color,soul}     % Пакет для выделения текста цветом
\sethlcolor{yellow}         % Цвет выделения по умолчанию (желтый)
\usepackage[autostyle]{csquotes}

\usepackage{graphicx}       % Для вставки графики
\graphicspath{{figures/}}   % Путь к папке с изображениями

\usepackage{dcolumn}        % Выравнивание столбцов таблиц по десятичной точке
\usepackage{bm}             % Жирные математические символы

\usepackage[mathlines]{lineno} % Нумерация строк (включая математические формулы)
\begin{document}

% Препринтная пометка (может быть удалена для финальной версии)
\preprint{APS/123-QED}

% Заголовок статьи (используем \\ для переноса строки)
\title{Influence of optical self-injection on statistical properties\\ of laser-pulse interference}

% Список авторов с аффилиациями
\author{Roman Shakhovoy}
\email{r.shakhovoy@goqrate.com}  % Контактный email
\affiliation{QRate, Moscow, Russia}
\affiliation{NTI Center for Quantum Communications, National University of Science and Technology MISIS, Moscow, Russia}

\author{Elizaveta Maksimova}
\affiliation{QRate, Moscow, Russia}

\author{Matvey Boltanskiy}
\affiliation{QRate, Moscow, Russia}

\author{Maxim Fadeev}
\affiliation{Russian Quantum Center, Skolkovo, Moscow, Russia}
\affiliation{NTI Center for Quantum Communications, National University of Science and Technology MISIS, Moscow, Russia}

% Дата (автоматически \today)
\date{\today}

% Аннотация
\begin{abstract}
	Pulsed optical self-injection markedly affects the emission characteristics of semiconductor lasers. In this work, we analyze its influence on the statistical properties of laser-pulse interference. We experimentally demonstrate that varying the arrival time of reflected optical pulses back into the laser cavity influences the phase-diffusion and induces an effect that, by analogy with external optical injection, we call phase locking. A comprehensive theoretical analysis of this phenomenon is also presented.
\end{abstract}

% Генерация титульной страницы
\maketitle

%%%%%%%%%%%%%%%%%%%%%%% INTRODUCTION %%%%%%%%%%%%%%%%%%%%%%%%%%%

\section{Introduction}\label{sec:introduction}
Lasers subject to optical feedback or, more generally, to optical self-injection (OSI) have been studied since the first demonstration of quantum generators (masers) more than 60 years ago \cite{Kleinman62,Rudd68}. OSI is particularly crucial for semiconductor laser diodes, whose inherently open-cavity structure makes them extremely sensitive to externally reflected or injected radiation \cite{Tartwijk95}.

The presence of OSI significantly complicates laser dynamics \cite{Lang80,Lenstra85,Henry86feedback,Petermann95} and can produce both undesirable effects, such as degraded direct modulation response or increased output noise \cite{Broom70,Ikushima78,Clarke91}, and beneficial ones, including spectral linewidth narrowing \cite{Bogatov73,Voumard77,Goldberg82,Petermann95}. At certain feedback coupling strengths, the system exhibits chaotic dynamics \cite{Mork92}; this behavior has been used to generate ultrawideband optical signals \cite{Zheng2010,Zhang2013}, has been proposed for LIDAR applications \cite{Chen2018,Cheng2018,Tsay2020}, and may prove useful in cryptography \cite{Mirasso2002}. Additionally, OSI has found intriguing applications in various contactless sensing techniques, particularly in optical feedback interferometry \cite{Liu2022}.

Considerable attention has been devoted to the influence of OSI on semiconductor laser characteristics, which has been extensively studied in the context of optical telecommunication systems \cite{Shikada88,Cartledge90,Clarke91,Langley93,Dellunde95,Homar97,Mohammad2011}. Typically, OSI effects are less pronounced in pulsed operation compared to continuous-wave regimes \cite{Fujita84,Petermann95}; moreover, signal distortions can be effectively suppressed if the reflected field reenters the cavity while the laser is off (between pump-current pulses) \cite{Ryan94}. However, under certain conditions, OSI can significantly degrade the output signal \cite{Cartledge90,Clarke91}, potentially increasing timing jitter \cite{Langley93}, generating additional spectral components \cite{Clarke91}, or closing the eye diagram \cite{Cartledge90}.

Significant research attention has focused on pulsed optical injection in the context of quantum key distribution (QKD) \cite{Yuan2016,Roberts2018,Paraiso2019,Lo2023simplified,Shakhovoy2021direct}. Of particular practical importance is the phase randomization effect of optical injection on laser pulses -- a phenomenon briefly discussed in \cite{Shakhovoy2021direct}. However, to our knowledge, the literature lacks detailed examination of OSI's impact on probabilistic properties of gain-switched lasers, particularly its effect on phase diffusion efficiency. This research gap is noteworthy as the problem holds relevance both for QKD implementations \cite{Kobayashi2014,Lo2007,Tang2013} and random number generation \cite{Quirce21,Valle2021,Shakhovoy2023}. Indeed, alongside the effects of chirp, jitter, and relaxation oscillations \cite{Shakhovoy2021}, the impact of OSI on interference can adversely affect optical random number generators that rely on inter-pulse phase randomness \cite{Jofre2011, Abellan2014, Yuan2014, Abellan2015, Marangon2018, Zhou2019, Shakhovoy2020}.

This work investigates the impact of OSI on the statistical properties of gain-switched laser-pulse interference. Section \ref{sec:experiment} presents experimental results demonstrating that varying the arrival time of reflected pulses affects not only the signal waveform but also the phase diffusion efficiency and may even induce \textit{phase locking} between laser pulses. To our knowledge, this phenomenon has not been previously reported in the literature. Sections \ref{sec:phaseSynchronization} and \ref{sec:simulations} provide an analysis of the phase locking mechanism arising from the pulsed OSI. Experimental findings and numerical simulations are discussed in Section\,\ref{sec:discussion}.

%%%%%%%%%%%%%%%%%%%%% EXPERIMENT %%%%%%%%%%%%%%%%%%%%%%%%%%%%%

\section{Experiment}\label{sec:experiment}

%====================== Experimental setup ===========================

\subsection{Experimental setup}\label{sec:experimentalSetup}
\begin{figure}[t]
\includegraphics[width=\columnwidth]{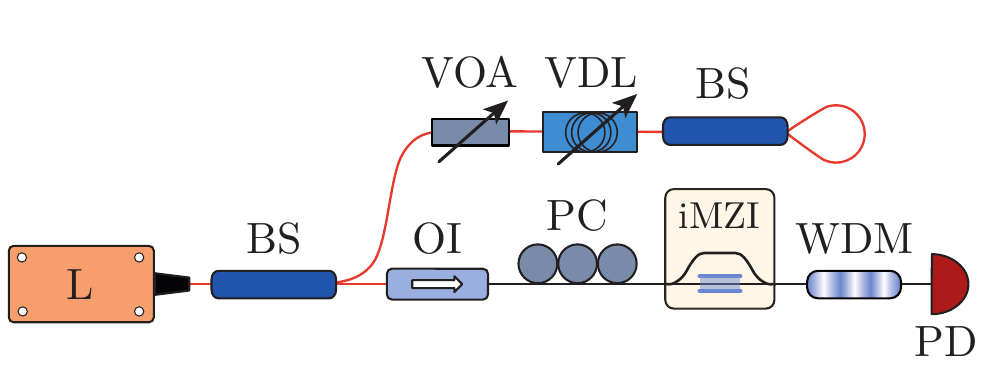}
\caption{\label{fig:experimentalSetup} Experimental setup: L -- laser, BS -- 50:50 beam splitter, VOA -- variable optical attenuator, VDL -- variable optical delay line, OI -- optical isolator, PC -- polarization controller, iMZI -- integrated Mach-Zehnder interferometer, WDM -- bandpass filter, PD -- photodetector.}
\end{figure}

The optical setup we used in the experiment is shown in Fig.\,\ref{fig:experimentalSetup}. A standard 1548.5\,nm distributed feedback (DFB) telecom laser (Nolatech 1550-DFB-14BF) was used as a light source. The laser output was split by a 50:50 fiber beam splitter (BS), with one port directed to a fiber loop mirror implementing optical feedback and another one -- to the Mach-Zehnder interferometer (MZI) with an 800\,ps delay line, where laser pulse interference occurred. A variable optical attenuator (Thorlabs V1550PA) and a delay line (IDIL Fibres Optics PM1550) were placed before the loop mirror to control feedback strength and external cavity length, respectively. The attenuation level was set to 0.5\,dB, corresponding to approximately 20\% of the original optical power that returned to the laser cavity. All optical components in the upper part of the schematic (colored red in Fig.\,\ref{fig:experimentalSetup}) were made from the polarization-maintaining (PM) fiber.

To prevent uncontrolled reflections, we placed an optical isolator at the beam splitter’s second output port. In front of the interferometer, we inserted a fiber polarization controller to align the light polarization with the MZI’s principal mode. A 100\,GHz DWDM filter (C36 channel) at the interferometer's output suppressed high-frequency spectral components associated with chirp. As demonstrated in \cite{Shakhovoy2021}, such spectral filtering reduces the influence of chirp, jitter, and relaxation oscillations on interference statistics.

The laser operated in gain-switched mode at 1.25\,GHz (800\,ps pulse interval) with an average output power of 2.7\,mW. The laser was driven by a custom QRate driver based on the Texas Instruments ONET1151L chip, injecting pump current as 400\,ps rectangular pulses at an 800\,ps repetition period. The high-frequency signal was generated by FPGA transceivers clocked at 156.25 MHz using a Silicon Labs frequency synthesizer, which itself was referenced to a 10 MHz signal from a high-stability clock oscillator. Laser pulses were detected using a Thorlabs PDA8GS photodetector (9.5 GHz bandwidth).

%======================== Experimental results =========================

\subsection{Experimental results}\label{sec:experimentalResults}
\begin{figure}[t]
\includegraphics[width=\columnwidth]{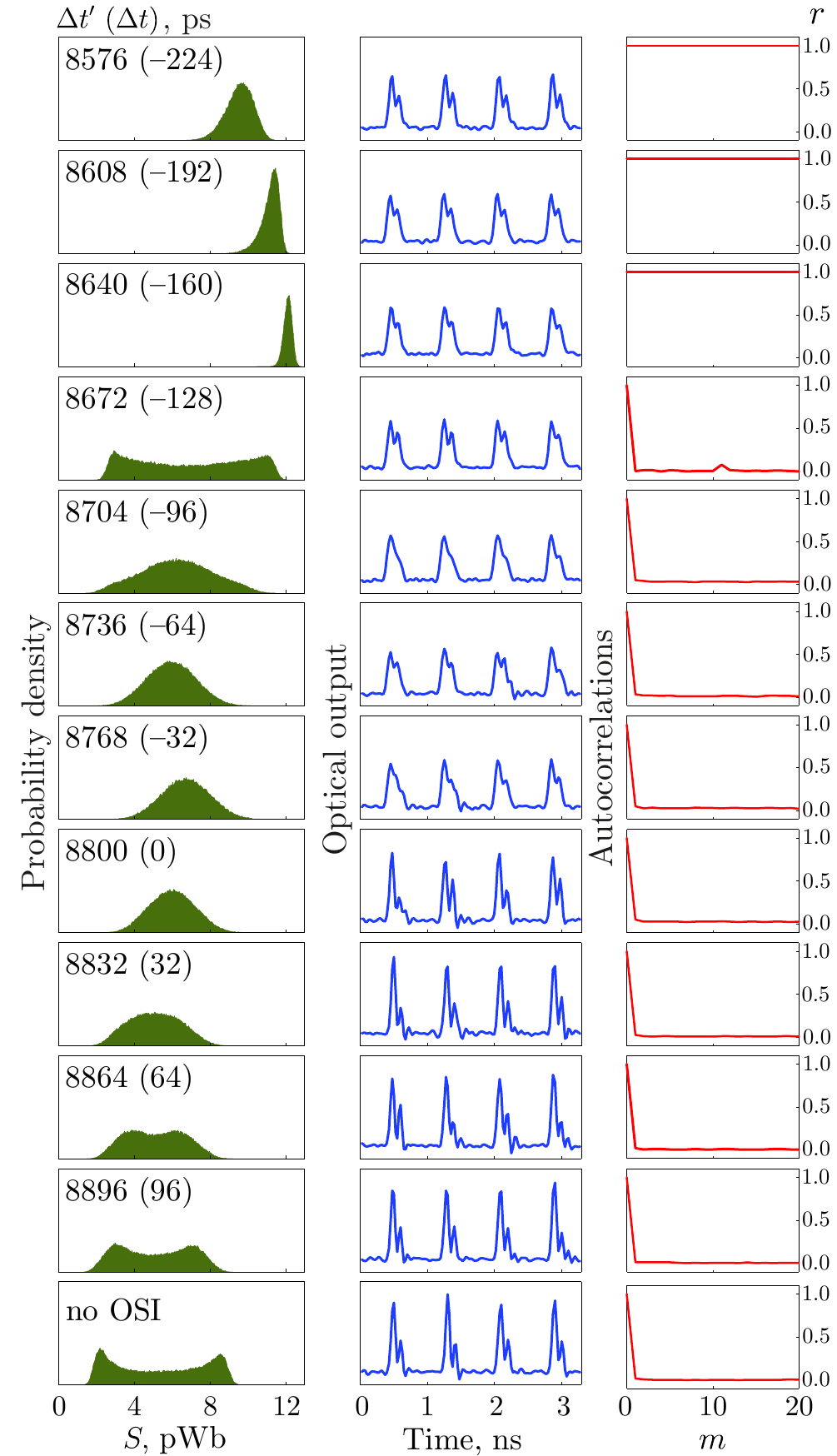}
\caption{\label{fig:experimentalResults} Experimental results. Left column: experimental probability density functions (histograms) of the interference signal for various external cavity lengths (expressed as delays between the reflected and newly generated lasers pulses). Middle column: laser pulses prior to the interferometer input. Right column: autocorrelation functions of the interference signal (pulse trains after the interferometer).}
\end{figure}

The experimental results are presented in Fig.\,\ref{fig:experimentalResults}. To demonstrate the effect of pulsed OSI on laser pulse shapes, the middle column displays short segments of pulse sequences acquired at different external cavity lengths \textit{prior to} interferometer input. (For these measurements, the photodetector was placed immediately after the optical isolator in Fig.\,\ref{fig:experimentalSetup}.) The right column presents autocorrelation functions of the pulsed signal \textit{after} the interferometer. Finally, the left column shows experimental probability density functions (histograms) of the interference signal for varying external cavity lengths.

We found it more practical to characterize the external cavity not by its physical length (in cm) but by the relative delay ${\Delta t=\Delta t'\,\text{mod} \, T}$ between the reflected and newly generated laser pulses, where $T$ is the pulse repetition period and $\Delta t'$ is the absolute delay corresponding to the round-trip time in the external cavity. The zero-delay reference (${\Delta t=0}$) was defined as the delay maximizing temporal overlap between reflected and new pulses. Direct experimental determination of optimal temporal overlap in our optical setup proved impossible; we instead inferred it from pulse shapes and interference signal distributions. Our analysis suggests that at good overlap (${\Delta t\approx0}$): (i) the relaxation oscillation peak should remain unsuppressed (confirmed numerically; see Fig.\,\ref{fig:simRes}), (ii) adjacent pulses should exhibit weak interference due to chaotic dynamics (evident in the ${\Delta t=0}$ statistics). Conversely, negative $\Delta t$ values should suppress relaxation oscillations. These criteria established our zero-delay reference. The absolute delay $\Delta t'$ can be independently determined from autocorrelation functions (discussed later). Both delays -- $\Delta t'$ and $\Delta t$ (in parentheses) -- are indicated in Fig.\,\ref{fig:experimentalResults}.

The autocorrelation functions in Fig.\,\ref{fig:experimentalResults} were computed as follows. First, each $i$-th pulse was assigned a value $S_i$, corresponding to its integrated area in picoweber units (V$\cdot$ps). The threshold was then calculated as ${\bar{S}=\frac{1}{2}\text{max}\,{S_i}}$, and the sequence was digitized according to the rule: ${S_i\to1}$ if ${S_i>\bar{S}}$; ${S_i\to -1}$ if ${S_i<\bar{S}}$. For the resulting sequence $x_0x_1...x_{N-1}$, where ${x_i\in\{-1,1\}}$, the pulse autocorrelation function was calculated using the following discrete form:
\begin{equation}
r(m)=\frac{1}{N}\sum_{i=0}^{N-m-1}x_ix_{i+m},\quad m=0,1,...,N-1.
\end{equation}

The autocorrelation functions for ${\Delta t \leq -160}$\,ps uniformly equal 1, indicating identical interference conditions for all pulse pairs (constructive in this case) with nearly constant phase differences. By analogy with optical injection, we term this phenomenon \textit{phase locking}. Notably, this effect is accompanied by partial suppression of the relaxation oscillation peak (see Fig.\,\ref{fig:simRes}), consistent with the similar behavior observed under optical injection \cite{Shakhovoybook}.

As $\Delta t$ increased from $-224$\,ps to $-160$\,ps, the relaxation oscillation suppression gradually weakened. At ${\Delta t = -128}$\,ps, an ${m=11}$ correlation emerged, which we attribute to the external cavity length ($\approx\!88$\,cm) corresponding to a round-trip time of $11T$, where ${T = 800}$\,ps is the pulse repetition period. No significant correlations were observed for ${\Delta t > -128}$\,ps.

As evident from Fig.\,\ref{fig:experimentalResults} (bottom row), without self-injection we observe the standard interference of laser pulses with random phases, where the signal's probability density function (PDF) exhibits two distinct maxima corresponding to destructive and constructive interference. The middle column (bottom row) displays gain-switched laser pulses prior to interferometer input, showing characteristic relaxation oscillation peaks. We emphasize that interference of such pulses would normally produce distorted statistics due to timing jitter \cite{Shakhovoy2021}. To recover the canonical two-peak PDF, we implemented a bandpass optical filter after the interferometer to suppress high-frequency spectral components.

For ${\Delta t > 0}$, when the reflected pulse re-enters the laser cavity after new pulse generation begins, phase locking vanishes, as evidenced by both the autocorrelation function and the PDF shape. This occurs because lasing initiates without reflected light, so each new pulse emerges with a random phase, which is typical for gain-switched operation. However, partial temporal overlap between returning and newly generated pulses induces chaotic dynamics that distort the interference statistics. Crucially, as pulse overlap decreases (increasing $\Delta t$), the distribution progressively converges toward the no-feedback case (bottom row of Fig.\,\ref{fig:experimentalResults}).

For ${-96 < \Delta t < 0}$\,ps, the pulse shape exhibits significant variations -- a signature of chaotic laser dynamics. This chaotic regime suppresses interference, causing the probability density function (PDF) to converge to Gaussian statistics.

At ${\Delta t = -160}$\,ps, pulse interference exhibits a narrow bell-shaped distribution, confirming efficient phase locking. While locking persists at ${\Delta t = -192}$\,ps and ${\Delta t = -224}$\,ps, the broadening PDF width indicates reduced efficiency compared to the ${\Delta t = -160}$\,ps case. We note that pronounced phase locking occurred only at small attenuation values (0\,--\,1\,dB); higher attenuation prevented stable locking or consistent pulse shapes. As shown later, the locking occurring at  ${\Delta t < -160}$\,ps  implies that phase differences between laser pulses are governed exclusively by external cavity round-trip phase accumulation and do not depend on their initial phases.

The ${\Delta t = -128}$\,ps case requires special consideration. Here, chaotic dynamics disappear, and the probability density resembles the no-self-injection case. However, the ${m=11}$ correlation reveals non-random phase differences between pulses. This correlation corresponds to an interference signal exhibiting an approximately repeating 11-pulse pattern, which is a consequence of optical feedback. Indeed, each of the first 11 pulses emitted at lasing onset determines the phase of its corresponding new pulse after completing the external cavity round-trip. This produces the repeating phase sequence $\varphi_1, \varphi_2, ..., \varphi_{11}, \varphi_1, \varphi_2, ..., \varphi_{11}, \varphi_1, ...$.

The observed behavior of the interference signal at ${\Delta t=-128}$\,ps can be explained as follows. Here, lasing starts in the presence of reflected light and this radiation seeds the phase of the field in the new laser pulse. As a result, the pulses are phase-coupled, despite the fact that the laser operates in the gain-switched mode. This is true, however, only in a very narrow interval of time delays in the vicinity of ${\Delta t=-128}$\,ps.

Such an informal description does not allow us to understand why, for some values of ${\Delta t<0}$, the phase differences between each pair of pulses become the same, i.\,e. why phase locking occurs. At first glance, such a result seems counterintuitive: we expected that for ${\Delta t<0}$ there would take place the situation that occurs at ${\Delta t=-128}$\,ps (see Fig.\,\ref{fig:experimentalResults}), i.\,e. an ${m=11}$ correlation would occur for each value of $\Delta t$ or interference would not be observed at all due to chaos, as for ${-96\le\Delta t\le0}$\,ps. However, as we have already found out, in a significant range of $\Delta t$ values (at least for ${-224\le\Delta t\le-160}$\,ps) phase locking is observed (${\varphi_{i+1}-\varphi_i=\varphi_{i+2}-\varphi_{i+1}}$ for all $i=1,2,...$), while the preservation of the values of the initially generated phases $\varphi_i$ is observed only in the vicinity of ${\Delta t=-128}$\,ps. In the next two sections, we will try to give a formal explanation of this effect. In section \ref{sec:phaseSynchronization}, a simplified model of phase locking is considered, which allows one to obtain useful analytical expressions, and in section \ref{sec:simulations}, the results of numerical simulations are presented.

%%%%%%%%%%%%%%%%%%%%% PHASE locking %%%%%%%%%%%%%%%%%%%%%%%%%%%%%

\section{Phase locking}\label{sec:phaseSynchronization}
The theoretical analysis of the experimental results obtained above is similar to the general analysis of the optical feedback, carried out by R.\,Lang and K.\,Kobayashi \cite{Lang80}, however, as far as we know, for pulsed self-injection such an analysis was carried out in the literature only numerically (see \cite{Cartledge90, Clarke91}). Here we will obtain an analytical expression for the phase difference between pulses in the case of pulsed OSI.

Let us note here that sometimes the terms \enquote{optical self-injection} and \enquote{optical feedback} are sometimes treated as distinct. The general argument is that in the presence of OSI, full-fledged feedback does not always manifest itself, e.\,g., it may be absent when reflected pulses return to the laser cavity between newly generated pulses. However, in our opinion, such a division is insignificant and only introduces confusion. Indeed, from a mathematical point of view, self-injection is always described by adding a time-shifted \textit{copy} of the signal to the equation, regardless of whether this leads to a dependence of laser generation on the prehistory or not. Formally, such a description corresponds to the introduction of feedback, although under certain conditions, it may indeed be very weak or not manifest itself at all.

According to experimental data, phase locking is observed when there is no chaos. In the experiment such a situation occurred only at ${\Delta t\le-160}$\,ps; however, in our simplified model, we will exclude chaos, which allows considering phase locking at ${\Delta t=0}$. Although this does not correspond to the experiment, such a simplification should not affect the physics of the locking process.

We start with the standard field rate equations for a semiconductor laser under OSI
\cite{Shakhovoybook,Lang80}:
\begin{equation}\label{eq:rateEquations}
\begin{split}
\dot{Q} &= (G - 1)\frac{Q}{\tau_\text{ph}} + C_\text{sp}\frac{N}{\tau_e} +\\
&+ 2\kappa\sqrt{Q(t)Q(t-\tau_\text{ex})} \cos[\Delta\varphi(t)- \varphi_\text{ex}]+F_Q, \\
\dot{\varphi} &= \frac{\alpha}{2\tau_\text{ph}}(G - 1) - \\
&-\kappa\sqrt{\frac{Q(t-\tau_\text{ex})}{Q(t)}} \sin[\Delta\varphi(t)-\varphi_\text{ex}]+F_\varphi,
\end{split}
\end{equation}
where
\begin{equation}
    \Delta\varphi(t)=\varphi(t) - \varphi(t-\tau_\text{ex}).
\end{equation}
Here $Q$ is the normalized intensity corresponding to the mean photon number in the cavity, $N$ is the carrier number, $G$ is the gain coefficient (we neglect gain non-linearity here), $\varphi$ is the field phase, $\varphi_\text{ex}$ is the additional phase acquired by a pulse during a round trip in the external cavity, $\tau_\text{ph}$ and $\tau_e$ are the photon and carrier lifetimes, respectively, $\kappa$ is the feedback-coupling coefficient, $\alpha$ is the linewidth enhancement factor, $F_Q$ and $F_\varphi$ are Langevin forces accounting for intensity and phase fluctuations, respectively. The system \eqref{eq:rateEquations} must be supplemented by a carrier rate equation, but this will not be needed for the present analysis.

For ideal overlap of the reflected and newly generated pulses (${\Delta t=0}$), we can set ${\dot{Q}(t)=f(t)Q(t)}$, where $f(t)$ is some continuous function of time; such an assumption is valid, in particular, in the case of Gaussian laser pulses: ${Q(t)=Q_0\exp[-t^2/(2w^2)]}$, for which ${\dot{Q}(t)=(-t/w^2)Q(t)}$. Given these assumptions, the rate equation for $Q$ can be written as the following algebraic equation:
\begin{equation}
\begin{split}
    G-1&=\tau_\text{ph}f(t)-2\kappa\tau_\text{ph}\cos(\Delta\varphi-\varphi_\text{ex})\\
    &-\frac{\tau_\text{ph}}{Q}\left(C_\text{sp}\frac{N}{\tau_e}+F_Q\right).
\end{split}
\end{equation}
Substituting it into the phase rate equation, we obtain:
\begin{equation}\label{eq:phaseEquation}
\begin{split}
\dot{\varphi}&=-\kappa[\alpha\cos(\Delta\varphi-\varphi_\text{ex})+\sin(\Delta\varphi-\varphi_\text{ex})]+\\ 
&+\frac{\alpha}{2}f(t)-\frac{\alpha}{2Q}\left(C_\text{sp}\frac{N}{\tau_e}+F_Q\right) + F_\varphi.
\end{split}
\end{equation}
The first term on the right-hand side of Eq.\,\eqref{eq:phaseEquation} represents the phase shift induced by the reflected field, the second term accounts for the phase change resulting from chirp, and the last two terms arise from spontaneous emission.

Further we will assume that ${\tau_\text{ex}=T}$, i.\,e. only the first laser pulse is generated without OSI. The second and subsequent pulses are generated in the presence of a reflected field. 

We will divide the change in the field phase during the first period (from 0 to $T$) into two time intervals. During the time interval when $Q$ is large and the contribution of spontaneous emission can be neglected, the phase dependence on time will be written as
\begin{equation}\label{eq:firstPulsePhase}
    \varphi(t)=\frac{\alpha}{2}\int_{0}^{t} f(t')\,dt'.
\end{equation}
During the remaining time, the evolution of the phase becomes random, so that the phase $\varphi(T)$ that the field acquires by the end of the period and which can thus be considered as the initial phase of the second pulse, is a random variable.

To determine the phase of the second pulse, we integrate Eq.~\eqref{eq:phaseEquation} from $T$ to $2T$, taking into account that the laser cavity now contains the reflected field. As before, we split the phase evolution into two intervals. During the time when $Q$ remains large, phase evolution is given by
\begin{equation}\label{eq:secondPulsePhase}
\begin{split}
    &\varphi(t)=\varphi(T)+\frac{\alpha}{2}\int_0^tf(t')dt'-\\
    &-\!\kappa\int\displaylimits_0^t\!\left\{\alpha\cos[\Delta\varphi(t')\!-\!\varphi_\text{ex}]\!+\!\sin[\Delta\varphi(t')\!-\!\varphi_\text{ex}]\right\}dt'.
\end{split}
\end{equation}
The phase evolution during the remainder of the second period is random, so the value of $\varphi(2T)$ is also random. To determine the phase difference between the first and second pulses, we subtract \eqref{eq:firstPulsePhase} from \eqref{eq:secondPulsePhase}, given that for the second pulse the phase \eqref{eq:firstPulsePhase} is $\varphi(t-T)$.  This yields
\begin{equation}
\begin{split}
    \Delta\varphi(t)&=\varphi(T)-\kappa\int_0^t\left\{\alpha\cos[\Delta\varphi(t')-\varphi_\text{ex}]\right.+\\
    &\quad\quad\quad\quad\quad\quad+\left.\sin[\Delta\varphi(t')-\varphi_\text{ex}]\right\}dt'.
\end{split}
\end{equation}
Differentiating this integral equation with respect to time, we obtain the following differential equation:
\begin{equation}\label{eq:diffEqForPhaseDifference}
    \frac{d\Delta\varphi}{dt}=-\kappa[\alpha\cos(\Delta\varphi-\varphi_\text{ex})+\sin(\Delta\varphi-\varphi_\text{ex})],
\end{equation}
whose solution for the initial condition ${\Delta\varphi(0)=\varphi(T)}$  is:
\begin{widetext}
\begin{align}\label{eq:phaseDifferenceSolution}
\Delta\varphi(t)=\varphi_\text{ex}+2\arctan\left\{\alpha^{-1}\left[1-\sqrt{1+\alpha^2}\tanh\left(\frac{\kappa t}{2}\sqrt{1+\alpha^2}+A\right)\right]\right\},
\end{align}
\end{widetext}
where
\begin{equation}
    A=\text{artanh}\left\{\frac{1+\alpha\tan\left[\frac{\varphi_\text{ex}-\varphi(T)}{2}\right]}{\sqrt{1+\alpha^2}}\right\}.
\end{equation}
Equation~\eqref{eq:phaseDifferenceSolution} shows that the phase difference $\Delta\varphi(t)$ evolves from
\begin{equation}
    \Delta\varphi(0)=-2\arctan\left\{\tan\left[\frac{\varphi_\text{ex}-\varphi(T)}{2}\right]\right\}
\end{equation}
to the asymptotic value
\begin{equation}\label{eq:finiteValue}
    \Delta\varphi(\infty)=\varphi_\text{ex}+2\arctan\left(\frac{1-\sqrt{1+\alpha^2}}{\alpha}\right)
\end{equation}
with a characteristic relaxation factor ${\exp(-\frac{1}{2}\kappa t\sqrt{1+\alpha^2})}$. In practice, the exponential term might be regarded as zero once it falls to $e^{-5}\approx\text{0,01}$; therefore, we assume that $\Delta\varphi$ reaches \eqref{eq:finiteValue} for $t>10/(\kappa\sqrt{1+\alpha^2})$. For a Fabry -- Perot laser, the coupling coefficient $\kappa$ can be estimated as \cite{Lang80}:
\begin{equation}
    \kappa=\frac{c(1-R)\sqrt{R_\text{ex}/R}}{2n_\text{g}L},
\end{equation}
where ${c\approx 3\cdot10^8}$\,m/s is the speed of light in vacuum, $L$ is the cavity length, $R$ is the power reflectivity of the laser facets, $R_{\text{ex}}$ is the reflectivity of the external mirror, and $n_{\text{g}}$ is the group index of the active layer. With typical values ${n_\text{g}=\text{3.5}}$, ${L=250}$\,\textmu m, and ${R=\text{0.3}}$, $\kappa$ ranges from several‐tens of GHz for ${R_\text{ex}=\text{0.1}}$ to about $200$\,GHz for ${R_\text{ex}>\text{0.9}}$. In the setup of Fig.\,\ref{fig:experimentalSetup}, ${R_\text{ex}\!\approx}$\,0.2\,--\,0.25, so ${\kappa\sim10^2}$\,ns$^{-1}$. With this value $\Delta\varphi$ reaches the limit defined by Eq.\,\eqref{eq:finiteValue} after roughly 20\,ps. For pulse durations of $\sim$300\,ps (as in Fig.\,\ref{fig:experimentalResults}) the interference outcome is governed mainly by $\varphi_\text{ex}$ (and by $\alpha$) and is essentially independent of the initial phase $\varphi(T)$. This is precisely the phenomenon of phase locking.

A similar analysis can be performed for all subsequent laser pulses. In particular, the phase of the third pulse (in the interval from $2T$ to $3T$) is again governed by
\eqref{eq:secondPulsePhase} now with the initial phase $\varphi(2T)$ in place of $\varphi(T)$). The phase of the preceding (second) pulse must also satisfy \eqref{eq:secondPulsePhase}; however, in the feedback term, one may substitute $\Delta\varphi(t)\approx\Delta\varphi(\infty)$. Since
\begin{equation}
\begin{split}
    &\alpha\cos[\Delta\varphi(\infty)-\varphi_\text{ex}]+\sin[\Delta\varphi(\infty)-\varphi_\text{ex}]=\\
    &=-\frac{\alpha}{\sqrt{1+\alpha^2}}+\frac{\alpha}{\sqrt{1+\alpha^2}}=0,
\end{split}
\end{equation}
this term vanishes, and we once again return to Eq.\,\eqref{eq:diffEqForPhaseDifference}. Hence, the phase difference between the second and third pulses is identical to that between the second and first ones, and so on.

Importantly, this result does not depend on the number of pulses that the laser emits before the reflected radiation appeared in the laser cavity: even if the external cavity \enquote{accommodates} a sufficiently long sequence of freely generated pulses, each of which has a random phase (as, in particular, happens in our experiment described in the section \ref{sec:experimentalResults}), all subsequent pulses will be phase-locked.

Thereby, the phase locking phenomenon demonstrated in Fig.\,\ref{fig:experimentalResults} at ${\Delta t<-128}$\,ps is explained by the fact that the phase difference between the pulses, according to \eqref{eq:finiteValue}, is determined by the phase incursion in the external cavity and does not depend on the initial phase of the newly generated pulse. We emphasize that this is a consequence of optical feedback in the sense that if a similar experiment on pulsed optical injection were carried out with an external laser operating in gain-switched mode, no similar effect would be observed.

Note, however, that the result obtained within our simplified model is applicable only to those experimental data in Fig.\,\ref{fig:experimentalResults}, where there is no chaos. In addition, the analysis we performed does not explain why phase locking does not occur at ${\Delta t=-128}$\,ps, where only a weak correlation associated with the length of the external cavity takes place. We attribute this discrepancy to the fact that the model does not take into account transient processes, for the analysis of which it is necessary to resort to numerical simulations.

%%%%%%%%%%%%%%%%%%%%% SIMULATIONS %%%%%%%%%%%%%%%%%%%%%%%%%%%%%

\section{Simulations}\label{sec:simulations}
To carry out the simulations, we adopt the standard model of stochastic field rate equation, now extended to include gain nonlinearity for greater generality. For computational convenience, we do not divide the complex electric field $E$ into $Q$ and $\varphi$, instead, we employ the following system \cite{Shakhovoybook}:
\begin{equation}\label{eq:RK}
    \begin{split}
    \dot{E} &= \frac{1}{2\tau_\text{ph}}[(G - 1)+i\alpha(G_L - 1)]E \\
        &+ \kappa E(t-\Delta t)e^{i\varphi_\text{ex}} + F_E,\\
        \dot{N} &= \frac{I}{e} - \frac{N}{\tau_e} - \frac{|E|^2}{\Gamma \tau_\text{ph}}G + F_N,
    \end{split}
\end{equation}
where $G_L = (N - N_0)/(N_\text{th} - N_0)$, and
\begin{equation}
    G = \frac{G_L}{\sqrt{1+2\gamma_Q|E|^2}}.
\end{equation} 
Here $N_\text{th}$ is the threshold carrier number, $N_0$ is the carrier number at transparency, $F_E$ and $F_N$ are Langevin forces that account for fluctuations of the optical field $E$ and carrier number $N$ respectively, $\Gamma$ is the optical confinement factor, and $\gamma_Q$ is the dimensionless gain-compression coefficient (all other symbols have the same meaning as in Eq.\,\eqref{eq:rateEquations}). Equation\,\eqref{eq:RK} implicitly assumes that the field $E$ is normalized such that $|E|^2$ equals the mean photon number $Q$ in the laser cavity. Finally, Langevin forces are written as follows: 
\begin{equation}
\begin{split}
    F_E&=\sqrt{\frac{C_\text{sp}N}{2\tau_e}}(\xi_1+i\xi_2),\\
    F_N&=-\sqrt{\frac{2C_\text{sp}NQ}{\tau_e}}\xi_1+\sqrt{\frac{2N}{\tau_e}}\xi_3,
\end{split}
\end{equation}
where $\xi_i$ are independent random variables that represent (normalized) white noise and satisfy the following relations:
\begin{equation}
    \langle\xi_i(t)\rangle=0,\quad\langle\xi_i(t)\xi_j(t')\rangle=\delta_{ij}\delta(t-t').
\end{equation}

The pump current in the simulations was set as a train of rectangular pulses, expressed as ${I(t)=I_\text{b}+I_\text{mod}(t)}$, where the bias current $I_\text{b}$ is time-independent and sets the minimum pump level, while the modulation current $I_\text{mod}(t)$ varies from 0 to $I_\text{p}$. Consequently, the pump current attains ${I_\text{max}=I_\text{b}+I_\text{p}}$. 

The laser and pump-current parameters used in the simulations are listed in Table \ref{tab:SimParameters}. The stochastic differential system \eqref{eq:RK} was integrated numerically with a fourth-order Runge–Kutta method \cite{Kloeden1994, gard1988introduction}.

\begin{table} 
	\caption{Simulation parameters.}
	\label{tab:SimParameters}
	\begin{center}		
		\tabcolsep=3pt		
		\begin{tabular}{lc}
			\hline
			\hline
			\rule{0mm}{3mm}
			Parameter & Value\\
			\hline
			\rule{0mm}{3mm}
			Bias current $I_\text{b}$,\,mA &6.0\\
			\rule{0mm}{2mm}
			Maximum pump current $I_\text{max}$,\,mA &150.0\\
			\rule{0mm}{3mm}
			  Carrier timelife $\tau_e$, ns &1.0\\
			\rule{0mm}{3mm}
			Photon timelife $\tau_\text{ph}$, ps &2.0\\
			\rule{0mm}{3mm}
			Pump current pulse width $w$, ns &0.4\\
			\rule{0mm}{3mm}
			Pulse repetition rate, GHz &1.25 \\
			\rule{0mm}{3mm}
			Confinement factor $\Gamma$ &0.12\\
			\rule{0mm}{3mm}
			Threshold carrier number $N_\text{th}$ &$5.5 \times 10^7$\\
			\rule{0mm}{3mm}
			Transparency level $N_{0}$ &$4.0\times10^7$\\
			\rule{0mm}{3mm}
			Spontaneous emissions fraction $C_\text{sp}$ &$10^{-6}$\\
			\rule{0mm}{3mm}
			Linewidth enhancement factor $\alpha$ &5.0\\
			\rule{0mm}{3mm}
			Gain compression factor $\gamma_Q$ &4.8$\times10^{-6}$\\
			\rule{0mm}{3mm}
			Feedback coupling coefficient $\kappa$, GHz &5.0 \\
			\hline
			\hline
		\end{tabular}
	\end{center}	
\end{table}

\begin{figure}[t]
	\includegraphics[width=\columnwidth]{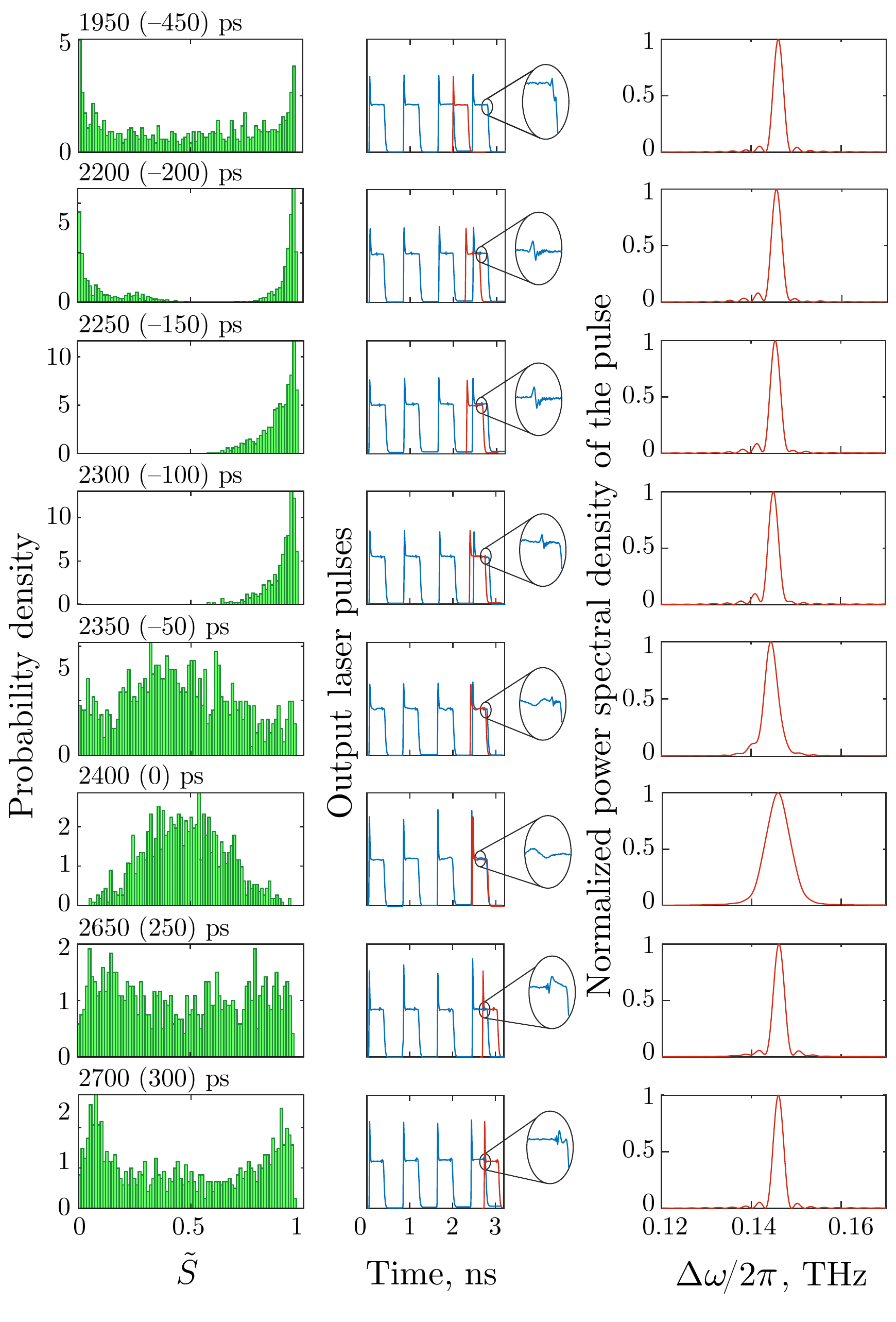}
	\caption{\label{fig:simRes} Simulation results. Left column:  histograms of the interference signal for various delay values. Center column: segments from the simulated pulse trains obtained by fourth-order Runge–Kutta integration of \eqref{eq:RK}. Right column: corresponding optical spectra (Fourier transforms of $E(t)$).}
\end{figure}

Figure\,\ref{fig:simRes} presents the simulation results that mirror the measurements in Fig.\,\ref{fig:experimentalResults}. For each delay value $\Delta t$, the model generated a train of $10^4$ pulses at a repetition rate of 1.25\,GHz. The interference signal was calculated as ${I(t)=|E(t)+E(t+T)|^2}$, where ${T=800}$\,ps, and the sequence of normalized interference signals $\tilde{S}_k$
\begin{equation}\label{eq:normalizedIntensity}
    \tilde{S}_k=\frac{\int\limits_{(k-1)T}^{kT}I(t)dt}{4\int\limits_0^T|E(t)|^2dt}
\end{equation}
was evaluated. $\tilde{S}_k$ values gives the histograms (left column of Fig.\,\ref{fig:simRes}). The right-hand column of Fig.\,\ref{fig:simRes} shows the theoretical spectra -- Fourier transforms of $E(t)$.

To speed up the computation, the external cavity in the model \enquote{accommodated} only 3 pulses (recall that in the experiment we had 11 pulses). This simplification is justified because, as argued earlier, phase locking is independent on the number of pulses that the laser emits before the reflected radiation appeared in the laser cavity. The only difference appears in the intermediate case, when occurs the correlation peak (as we observed at ${\Delta t=-128}$\,ps; see Fig.\,\ref{fig:experimentalResults}). In simulations, we would obtain the ${m=3}$ correlation, rather than ${m=11}$.

To demonstrate the pulse shape for different values of $\Delta t$, we show the segments of the pulse sequences in the middle column of Fig.\,\ref{fig:simRes}. Note that the time axis in the plots is shifted to the right, so that all pulses shown in the figure were generated in the presence of OSI. Note also that the finite bandwidth of the photodetector (or oscilloscope), which leads to signal broadening in the experiment, was not taken into account in the simulations. We also did not select the laser parameters with high accuracy in order to obtain good agreement between the experimental and theoretical laser pulse shapes, since this has almost no effect on the result. For clarity, one of the reflected pulses is shown in red. Finally, oval insets magnify features caused by the reflected field.

\begin{figure}[t]
	\includegraphics[width=\columnwidth]{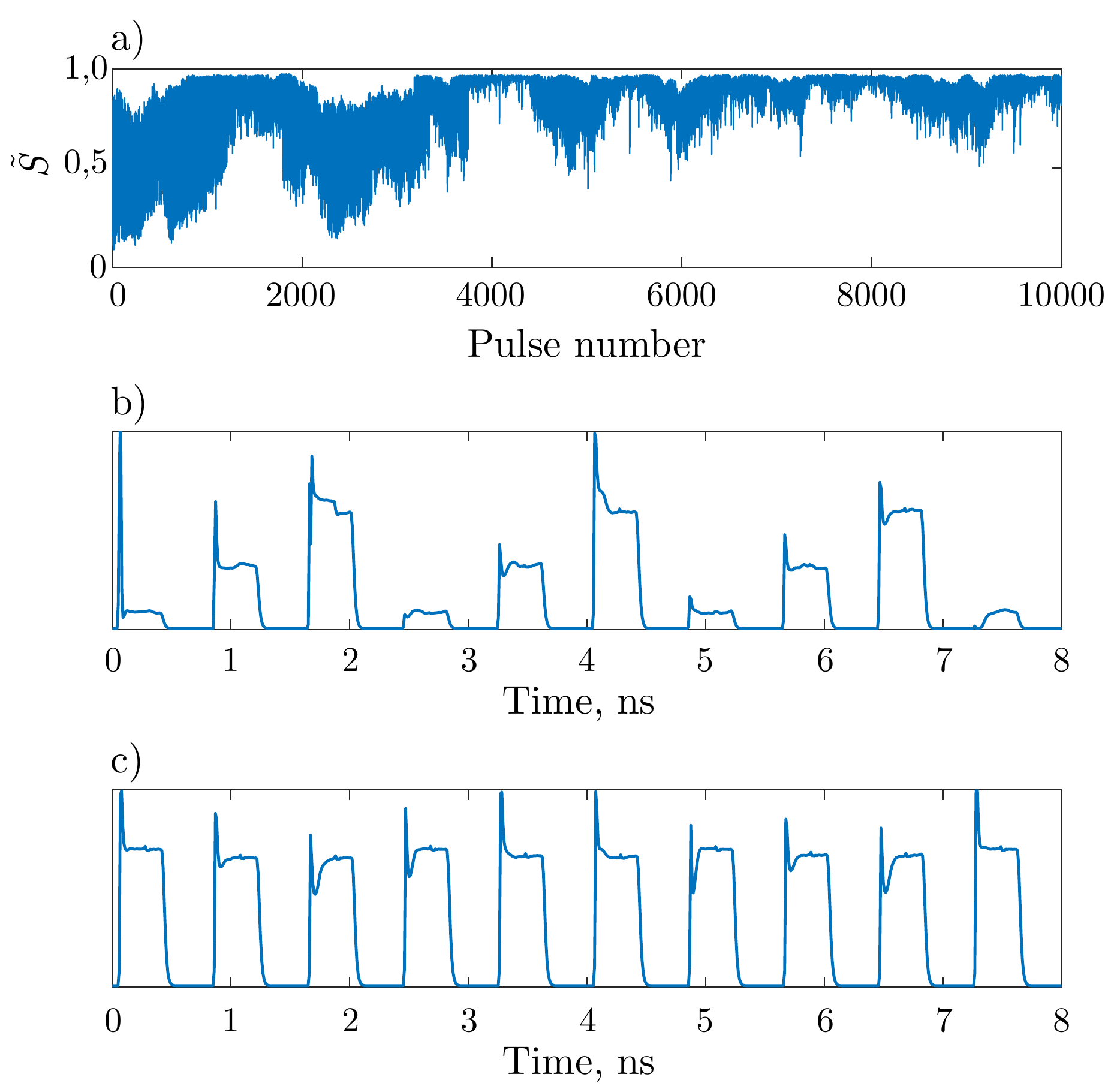}
	\caption{\label{fig:pulsesStab} (a) Normalized interference signal defined by Eq.\,\eqref{eq:normalizedIntensity} at ${\Delta t=-\text{150}}$\,ps. (b) First 10 pulses from the train of 10000 pulses. (c) Last 10 pulses from this train.}
\end{figure}

At ${\Delta t=-450}$\,ps, the reflected pulses return \textit{between} the newly generated pulses. Figure\,\ref{fig:simRes} shows that only the trailing edge of each new pulse \enquote{feels} the first relaxation spike of the reflected signal (see the oval inset). Phase locking does not occur in this case, since each new laser pulse appears in an almost empty cavity and therefore has a random phase. This is clearly seen from the characteristic shape of the histograms. If lasing starts in the presence of reflected radiation, as happens, e.\,g., at ${\Delta t=-150}$ and $-100$\,ps, then, according to the analysis performed in section \ref{sec:phaseSynchronization}, phase locking occurs. It is important, however, that it does not appear instantly, but is preceded by a fairly long transient process, which is demonstrated in Fig.\,\ref{fig:pulsesStab}.

Figure\,\ref{fig:pulsesStab}(a) shows the normalized power of the interference signal at ${\Delta t=-150}$\,ps for $10^4$ pulses. The values of $\tilde{S}_k$ initially have a large spread, but then the range of normalized power values becomes smaller, which is perceived in the experiment as phase locking. Figure\,\ref{fig:pulsesStab}(b) shows the first ten pulses of the same interference signal, where a repeating pattern of three pulses is clearly visible. This pattern arises due to the fact that the first three laser pulses generated in the absence of OSI have random phases $\varphi_1$, $\varphi_2$, $\varphi_3$. After several thousand pulses, the values of these phases converge, and by the end of the train they become approximately the same. This convergence is evident from Fig.\,\ref{fig:pulsesStab}(c), which displays the last ten pulses of the signal shown in panel (a).

Thereby, a key feature of phase locking not captured by our simplified analytical model from the preceding section are the long transients that precede the locking.  Interestingly, for some delay values $\Delta t$ the transient never terminates, and the repeating pattern persists, producing a sharp peak in the autocorrelation function. Such a situation occurs in Fig.\,\ref{fig:simRes} at ${\Delta t=-200}$\,ps (the autocorrelation with its peak at $m=3$ is not shown).  A similar behavior was observed experimentally (see Fig.\,\ref{fig:experimentalResults} at $\Delta t=-128\,$ps). The reason why the transient process does not terminate is not yet clear to us; it is presumably related to the emergence of chaos. Note that the transients illustrated in Fig.\,\ref{fig:pulsesStab} were difficult to observe in the laboratory, because the initial laser‐pulse generation was accompanied by additional long transients in the driver circuitry as well as thermal settling in the diode itself (see \cite{Shakhovoy2023synch} for details).

The influence of chaos is most pronounced in Fig.\,\ref{fig:simRes} at ${\Delta t=-50}$ and 0\,ps, where a significant overlap of the reflected and newly generated laser pulses occurs. At ${\Delta t=0}$ the histograms of the interference signal becomes Gaussian, indicating poor interference. Also note the shape of the signal shown in the oval insets at ${\Delta t=-50}$ and 0\,ps: instead of a flat \enquote{shelf}, chaotic oscillations are observed. These oscillations are accompanied by a chaotic phase change, which is reflected in a significant broadening of the spectrum (see the corresponding spectra to the right of the pulses in Fig.\,\ref{fig:simRes}).

At ${\Delta t=250}$ and 300\,ps, when the reflected pulses return to the laser cavity long after the onset of lasing, no phase locking is observed, and the influence of chaos that occurs in the overlap region of the reflected and new laser pulses is weaker. Interference quality therefore improves and the spectrum narrows.

All simulation results of numerical simulation agree very well with the experiment. Note, however, that the experimental statistics look \enquote{smoother} since they were obtained on a larger amount of data. A further difference appears between the theoretical statistics at ${\Delta t=-200}$\,ps in Fig.\,\ref{fig:simRes} and the experimental statistics at ${\Delta t=-128}$\,ps in Fig.\,\ref{fig:experimentalResults}. Both statistics correspond to the regime without chaos and without phase locking. Here the PDF is the sum of independent statistics from each of the pulses in the repeating interference pattern. In the experiment there were ${m=11}$ such independent statistics, and in the simulations -- only ${m=3}$, which is why the resulting PDFs have different shapes.

%%%%%%%%%%%%%%%%%%%%% DISCUSSION %%%%%%%%%%%%%%%%%%%%%%%%%%%%%

\section{Discussion}\label{sec:discussion}
\begin{figure}[t]
	\includegraphics[width=\columnwidth]{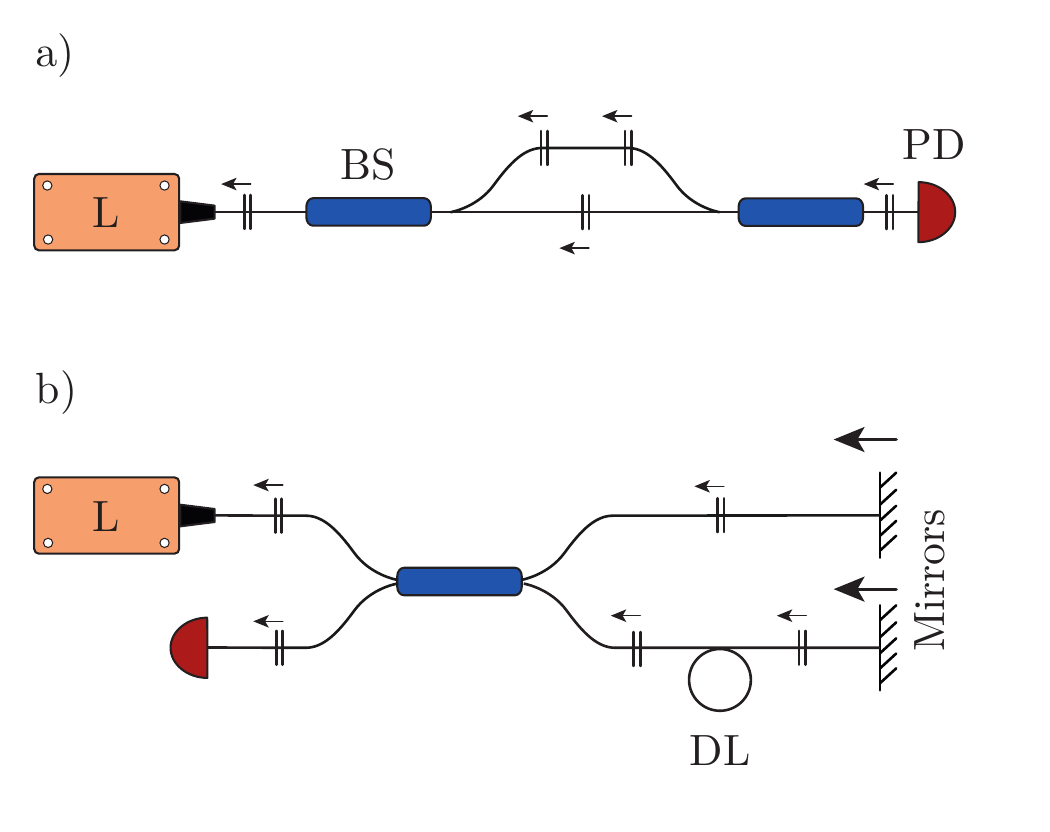}
	\caption{\label{fig:schematicsWithInterferometers} 
		Possible fiber-optic circuits for Mach–Zehnder (a) and Michelson (b) interferometers: DL -- delay line (the rest notation is as in Fig.\,\ref{fig:experimentalSetup}). Short arrows represent the reflections on the connectors.
	}
\end{figure}

The problem of uncontrolled OSI arises whenever emission from a semiconductor laser is coupled into an optical fiber.  It can be effectively mitigated by a spherical lens formed on the fiber endface \cite{Byrne91}, which yields a reflectivity of \(-40\) to \(-50\)\,dB, so feedback is usually neglected. However, fiber-optic circuits are often built from discrete components, introducing numerous connector interfaces that give rise to parasitic reflections. Figure\,\ref{fig:schematicsWithInterferometers} illustrates the possible reflections in Mach -- Zehnder and Michelson interferometer circuits: short arrows denote connector reflections, and long arrows denote mirror reflections.  In a Michelson interferometer back-illumination into the laser -- and hence OSI -- is much stronger than in a Mach -- Zehnder, so additional countermeasures (e.\,g., an optical isolator) are generally required.  Nevertheless, OSI remains an issue in discrete-component fiber-optic systems and must be taken into account.

The demonstrated effect of phase locking may have negative consequences in some applications that employ gain-switched semiconductor lasers in the interferometer configurations similar to that in Fig.\,\ref{fig:schematicsWithInterferometers}.  In particular, it can undermine the security of quantum key distribution protocols \cite{Kobayashi2014,Lo2007} or enable side-channel attacks \cite{Sun2015,Huang2019,Pang2020,Lovic2023}.  Moreover, the induced correlations in the interference signal can degrade the randomness of binary sequences generated by optical random-number generators based on laser-pulse interference \cite{Jofre2011,Abellan2014,Yuan2014,Abellan2015,Marangon2018,Zhou2019,Shakhovoy2020}. For instance, the correlation observed at ${\Delta t=-128}$\,ps in Fig.\,\ref{fig:experimentalResults} could impair the statistical quality of the output if left unaddressed. This work was in part motivated by the desire to eliminate the optical isolator from such random number generator circuits for simplicity and cost reduction; our findings indicate, however, that an isolator remains essential.

We emphasize once again that the phase difference $\Delta\varphi(\infty)$ established as a result of phase locking does not depend on the initial (random) phase of the laser pulse, but only on the phase accumulated in the external cavity. This makes the interference sensitive to changes in the external cavity, in particular, to changes in its length. In our experiment, the external cavity was just a segment of fiber, so the interference outcome drifted over time due to temperature fluctuations, remaining stable for only a few seconds. To acquire the statistics in Fig.\,\ref{fig:experimentalResults}, we used pulse trains of 100\,\textmu s duration so that the effect of temperature on the PDF's shape could be neglected.

Note also an interesting feature related to the amplitude of the interference signal in Fig.\,\ref{fig:experimentalResults}. Comparison of the PDFs shows that the maximum signal level $S$ for constructive interference without OSI reaches approximately 9\,pWb, whereas in the presence of phase locking the maximum signal level (at ${\Delta t=-160}$\,ps) reaches 12\,pWb. This feature is due to the fact that we used spectral filtering before detection: without OSI laser pulses are characterized by a significant chirp, which is cut off by the DWDM filter, so the part of the spectral power is lost, whereas with OSI chirp decreases and more optical power passes through the DWDM filter to the detector.

%%%%%%%%%%%%%%%%%%%%% CONCLUSION %%%%%%%%%%%%%%%%%%%%%%%%%%%%%

\section{Conclusion}\label{sec:conclusion}
We have demonstrated how pulsed optical self-injection influences the statistical properties of laser-pulse interference, showing that varying the arrival time of the reflected pulses affects the efficiency of phase diffusion and can induce phase locking -- an effect not previously reported in the literature. For sufficiently accurate temporal overlap of laser pulses, the equation for the phase (or more precisely for the phase difference $\Delta\varphi$ between adjacent pulses) admits an analytical solution (provided that there is no chaos). This solution predicts that $\Delta\varphi$ approaches a stationary value that does not depend on the initial (random) pulse phase,  but is determined by the phase incursion that the laser pulse acquires when traversing the external cavity. The simplified analysis, however, neglects the long transients preceding the establishment of the stationary value of $\Delta\varphi$, which can last for several thousand pulse repetition periods. 

The phase-locking effect reported here must be taken into account when assessing the secrecy of quantum-key-distribution protocols and when analyzing the statistical properties of binary random sequences, whenever a fraction of the reflected power is able to re-enter the laser cavity.

\bibliography{selfInjection}

\end{document}